\documentclass[11pt]{iopart}
\usepackage{iopams}
\usepackage{txfonts}
\usepackage{graphicx}

\newcommand{\halpha}{H$\alpha$}
\newcommand{\arcsec}{$^{\prime\prime}$}

\newcommand{\msun}{M$_\odot$}

\begin{document}

\article[Molecular gas in Tidal Dwarf Galaxies]{}{Molecular gas in Tidal Dwarf Galaxies: Exploring the conditions for star formation}

\author{U.\  Lisenfeld$^1$, F.\ Bournaud$^2$, E. Brinks$^3$, P.-A.\ Duc$^2$}

\address{$^1$ Dept.\ de F\'\i sica Te\'orica y del Cosmos, Universidad de Granada, Spain\\
$^2$ Laboratoire AIM, CEA/DSM -- CNRS --
Universit\'e Paris Diderot, DAPNIA/Service dÕAstrophysique, CEA/Saclay, F-91191 Gif-sur-Yvette
Cedex, France\\
$^3$ Centre for Astrophysics Research, University of Hertfordshire, College Lane, Hatfield AL10 9AB, UK
}
\ead{\mailto{ute@ugr.es,frederic.bournaud@cea.fr,paduc@cea.fr,e.brinks@herts.ac.uk}}

\begin{abstract}
Tidal Dwarf Galaxies (TDGs), produced from material expelled
in galactic interactions, are well--suited  to test the laws of star formation (SF) 
due to their simple structure, high metallicity -- 
making CO a reliable tracer of the molecular gas content -- and
recent SF.
Here, we study the conditions for the onset of SF and for the rate at which SF proceeds once
above a threshold in a small sample of TDGs.
We use data  for the gas (atomic and molecular) surface  density and SF rate per
area to test the laws of SF found for spiral and dwarf galaxies in this more
extreme environment.
We find in general a good agreement with the Schmidt law found for the
total gas \cite{kennicutt98} and for the molecular gas \cite{bigiel} but note that higher
resolution CO observations are necessary to clarify some possible discrepancies.
We find, down to a scale of $\sim$1 kpc, in general a good agreement between the
peaks of SF and of the molecular gas, but also find in some objects 
surprisingly large quantities of molecular gas at places where no SF is 
occuring.\ A high column
density of molecular gas is therefore not a sufficient condition for the onset of SF. 
We find that the kinematical properties of the gas are also relevant: in two objects
our observations showed that SF only occured in regions with a narrow line width. 
\end{abstract}


\section{Introduction \label{S1}}

A major  challenge in astrophysics is  understanding the
process of star formation (SF) in galaxies and in particular
to find out (i) what determines the threshold
for star formation, i.e., what are the sufficient and necessary conditions for SF to 
take place and (ii) what determines, once above this threshold,
the rate of SF. This knowledge is not only useful in itself, but forms
a crucial ingredient for the modeling of galaxy formation and evolution 
(e.g., Schaye \& Dalla Vecchia \cite{schaye08}).

Tidal Dwarf Galaxies (TDGs) are small galaxies which are currently in the
process of formation. They are forming 
from material ejected from the disks of spiral galaxies through  galactic 
collisions.
Their properties are very similar to those of classical dwarf irregulars
and blue compact dwarf galaxies, except for their metallicities which
are higher and lie in a narrow range of 12+log(O/H)$\approx 8.4 - 8.6$
\cite{duc00}. These metallicities
are typical of the material found in the outer
spiral disks of the interacting galaxies and which is
most easily lost to form TDGs.

TDGs are well suited  to test the laws of  SF because of various
reasons: In the same way as "classical" dwarf galaxies (i.e., dwarf galaxies
not obviously formed from an interaction) the lack of differential
rotation  and, more importantly, the absence of
prominent density waves make shear and
centrifugal forces less important and reduce in this way the number of parameters that
affect SF. Furthermore, 
TDGs lack an old
stellar population, because they are made  almost entirely from a gaseous medium  with none or
only few old stars coming from the parent galaxies,
 so that their gravitational potential is largely dominated by
the gas distribution alone.
Additionally, an  advantage of TDGs with respect to classical dwarf galaxies is
their higher metallicity which makes CO a good tracer of the molecular gas
and thus allows us to directly study the relation between molecular gas and SF.

The goal of this contribution is to discuss the relation between 
SF and molecular gas in TDGs. To this aim, the existing observations and their
basic results are reviewed and compared to different theories describing
 the laws of SF and conditions for the onset of SF.

\section{Observations of the molecular gas}

The first detection of molecular gas in a sample of TDGs \cite{braine00,braine01} revealed 
in all objects  a good  
spatial overlap between the areas where molecular gas is detected, 
the sites of SF and the peak of the
atomic gas. Furthermore, the kinematical properties of the atomic and molecular
gas are identical, i.e., both lines show the same central velocity and line width.

These single--dish observations were taken with only a modest  spatial resolution
(mostly  about 22\arcsec , corresponding to a linear resolution between 2 and 14 kpc 
at the distances  of the objects, between 22 and 135 Mpc). Nevertheless, they allowed us to
determine the molecular-to-atomic gas mass ratio 
and the molecular star formation efficieny (SFE, defined here
as the SF rate devided by the molecular gas mass) and
compare them to those of spiral galaxies. The molecular gas mass
in TDGs was calculated adopting a Galactic conversion factor ($N_{\rm H2}/I_{\rm CO} 
= 2\times 10^{20}$ cm$^{-2}$(K km s$^{-1}$)$^{-1}$), the same factor adopted in
the present contribution.
Contrary to classical dwarf
galaxies, TDGs show both a molecular gas fraction and a SFE 
in the range of spiral galaxies \cite{braine01,lisenfeld02}. The authors concluded that
 (i) the CO-to-H$_2$ conversion factor
is  indeed similar to the Galactic value and not,
as in classical dwarf galaxies, much lower and (ii)  SF proceeds in a normal 
way in TDGs, similar to that occurring in spiral galaxies.

Since then, a few objects have been observed at higher spatial resolution.
Interferometric observations have been carried out in two objects, the northern TDG, Arp 245N,
in   the interacting system NGC 2992/3 (Arp 245) \cite{brinks} and the
TDG SQ B in Stephan's Quintett \cite{lisenfeld04}.
These observations confirmed at a higher spatial resolution that 
 the peak of the molecular gas coincides reasonably well with recent SF, traced by \halpha ,
 although at the highest spatial resolution,
 in Arp 245N a marginally resolved displacement of about  3-4\arcsec (400 -- 600 pc)
 was found between both peaks. 
 The comparison of the total flux of the interferometric
and single-dish observations showed that a large fraction (25\% for 
Arp 245N and 50\% for SQ B)
of the flux is missing in the interferometric observations, indicating
the presence of a considerable amount of  smoothly distributed molecular
gas. 

Wide-spread molecular gas was also found in two 
nearby objects where the single-dish observations provided enough
linear resolution to create maps. One object is an  old TDG
in the Virgo Cluster, VCC 2062 \cite{duc07} and the other is a 
potential TDG in the interacting system Arp 94, J1023+1952
\cite{lisenfeld08}.
In both objects, the star--forming region sits within a large HI cloud. 
Surprisingly, 
molecular gas was not only found to coincide with
SF regions, but also, in  large quantities,
at places where no SF is ocurring.
The distribution of the molecular gas follows, although not 
at  a strictly  constant ratio, that of the atomic gas and
the kinematical properties of HI and H$_2$ are  similar.
Neither the column density of the molecular nor the column density
of the total (atomic plus molecular) gas seems to be a relevant parameter to explain why 
SF  is present in only one region of the gas clouds,  and  
in particular a lack of
molecular gas is not  the reason for the absence  of SF over a large
extent of the cloud.
A similar result was obtained by \cite{maybhate} who found in a 
survey of tidal tails that a high column density
($\log N_{\rm HI} > $ 20.6 cm$^{-2}$) is a necessary but not sufficient condition for SF to take place.

We found, however, a kinematical difference in the gas properties:
the line width in the SF region is narrower
(FWHM $\sim$ 30-70 km s$^{-1}$ for J1023+1952 and FWHM $\sim$ 20 km s$^{-1}$ for
VCC 2062) than in the region without
SF where the line width is about a factor of 2 higher. This  indicates that dynamically cold gas is a necessary condition
for SF to take place.

In summary, we found the following common features in the properties
of the molecular gas and SF in TDGs:

\begin{itemize}

\item There is  a close relation between
atomic and molecular gas, both spatially and kinematically.
The molecular-to-atomic gas mass ratio  is similar to that of spiral galaxies.

\item In SF regions, molecular gas is present and the SFE in TDGs is similar 
to that of spiral galaxies.

\item The distribution of the molecular gas is {\it not} limited to the SF region, and 
abundant molecular gas  was found over a large extent of the HI distribution.

\item The column density of the molecular gas and of the total gas is not
the key parameter for the presence of SF. The kinematical properties of the
gas play an important role as well in the sense that SF only takes place
in dynamically cold gas.

\end{itemize}

\section{Testing the laws of  star formation}

In this section we  compare the observations to existing theories
for SF. We start with a review of the most common theories.

\subsection{Theories of star formation}

In a classical paper, \cite{toomre} derived a prescription for the
stability of a gas disk against collapse and star formation, 
taking into account gravitation on one hand and 
internal pressure and galactic rotation as stabilizing 
processes on the other. This Toomre--criterion has 
shown to be a reasonably good predictor of the SF threshold 
(e.g., Kennicutt  \cite{kennicutt98}),
especially when the stellar mass component is properly taken into account
to describe the gravitational field \cite{jog}.
Since in  dwarf galaxies differential rotation is not very important,
shear has been suggested as an alternative mechanism to prevent gravitational
collapse \cite{elmegreen87, elmegreen91,elmegreen93} and has been found to be a better description
for the onset of SF in dwarf galaxies \cite{hunter}.
However, SF has also been found in regions which are globally
well below the SF threshold \cite{ferguson, braine07}.
This is explained in general terms by the fact that the Toomre parameter 
does not  represent a strict lower threshold in the sense that local
parameters like the magnetic field, turbulence and variations in the
local velocity dispersion can allow SF to take place even below the global 
threshold \cite{schaye04}.  An alternative explanation is that the mass of star clusters
in the outer disk environment is decreasing, leading to an reduced  \halpha\ emission 
for a given SF rate (SFR) \cite{pflamm}.
A strict lower limit for SF is expected to be set by the minimum density
providing the necessary pressure to maintain cool atomic gas \cite{elmegreen02,
schaye08}. This lower limit is expected to be around 3 - 10 \msun\,pc$^{-2}$
but will depend on parameters like the radiation field and
the metallicity. The exact relation between the threshold and
these parameters is, however, still a matter of debate. 

Once supercritical conditions are reached and SF starts,
its rate is expected to depend on the gas density and the characteristic
time scale for cloud collapse.  Together this leads to the prediction
by Schmidt \cite{schmidt59},  who suggested
that the SFR is proportional to the gas volume density to the power $n$, 
where $n=1.5$ (note that for a constant scaleheight gas disk, the
same relation holds between the SFR density in the disk and gas surface 
density). Several studies have found $n$ to lie between
1 and 2. \cite{kennicutt98} determined $n$ to be close to $1.40 \pm 0.15$ 
over 5 orders of magnitude.
The precise value of $n$ depends on many factors which complicate its 
interpretation. For example, although one would expect total, i.e., atomic 
plus molecular hydrogen, gas surface density to drive the SFR,
Wong \& Blitz \cite{wong} argue that in fact it is only the H$_2$ which is
involved. This is now being confirmed on a much larger sample using
THINGS (The HI Nearby Galaxy Survey) data \cite{bigiel,leroy} .

Different prescriptions have been proposed to explain the SFE, such as
the Toomre criterion, a
constant column density threshold \cite{schaye04}, Silk's suggestion of 
star formation being dependent on a dynamical time scale \cite{silk},
as well as  the model  by 
Blitz \& Rosolowsky \cite{blitz} who suggest that the transformation 
between atomic and molecular gas, and thus the relation between
gas and SF, is solely determined by hydrostatic pressure.
Additionally, large scale turbulence (i.e., the velocity
dispersion between molecular clouds)  might play a role.
Elmegreen \cite{elmegreen02} suggests that large-scale turbulence can play a double role:
it can  trigger SF by compression of preexisting gas clouds, but
it can also inhibit SF if the motions continously force the
gas to break up to pieces that are smaller than
a thermal Jean mass.
Using THINGS data,  Leroy et al. \cite{leroy}
tested a number of these prescriptions in dwarf and
spiral galaxies. Of all prescriptions investigated, the model by
Blitz \& Rosolowsky  \cite{blitz} is
the one which performed best throughout, although 
none of the other theories can be ruled out decisively,
even though they are based on very different physical assumptions.

\subsection{Comparison of theory and observations}

In Lisenfeld et al. \cite{lisenfeld02} we showed that the
SFR and gas column density of the sample of TDGs observed up to that 
date and at a modest resolution (22\arcsec for the CO)
 followed the Schmidt-law found by Kennicutt \cite{kennicutt98} reasonably well.
Here, we want to test this again for those  objects for which in the
meanwhile spatially resolved observations have become available.
Additionally,  we compare
our data to more recent results of Leroy et al. \cite{leroy} and Bigiel et al. \cite{bigiel},
based on the THINGS survey.

We determined the SFR from the extinction corrected \halpha\ luminosity
following the prescription given in Bigiel et al.  \cite{bigiel}

\begin{equation}
{\rm SFR} [{\rm M}_\odot {\rm yr}^{-1}] = 5.3 \times 10^{-42} L_{\rm H\alpha} [{\rm erg \, s^{-1}}]
\end{equation}

\begin{table}
\caption{SFR surface density and   gas surface densities}
\begin{tabular}{lccccccc}
 
\hline 

\hline

Name & Distance & resolution$^{a}$ & $\Sigma_{\rm HI}$ & $ \Sigma_{\rm H2}$ & $ \Sigma_{\rm SFR}$ &$\Sigma_{\rm H2}/\Sigma_{\rm HI+H2}$ &   SFE(H$_2$)$^{b}$\\
 & [Mpc] & [\arcsec/kpc]  & \multicolumn{2}{c}{[M$_\odot {\rm pc}^{-2}$] } & [$10^{-3}$M$_\odot {\rm yr}^{-1}{\rm kpc}^{-2}$]  &  & [10$^{-10}$yr$^{-1}$] \\
\hline
VCC 2062 & 17 & 22\arcsec/1.8kpc & 7.1 & 2.7   & 0.2 -- 1.1  & 0.28 & 0.9 -- 4.1 \\
Arp245N  & 31& $\sim$ 4\arcsec/0.6 kpc & 18.2$^{c}$  & 10.0 & 24  & 0.35 & 24  \\
J1023+1952 & 20.4    & 22\arcsec/2.2kpc & 9.5   & 4.7    & 4.2 -- 8.4  & 0.33 & 9 --18\\
SQ B      & 85 & $\sim$ 4\arcsec/1.6kpc & 5.5$^{c}$   & 12    & 17  & 0.68 & 14 \\
\hline
\end{tabular}

$^{a}$ Resolution of the CO. The resolution of the  HI data is about 20\arcsec\ for all objects.
 
 $^{b}$ The molecular star formation efficiency, SFE(H$_2$),  defined as the SFR per molecular gas mass. The range of values given for VCC2062 and J1023+1952 is due to different areas
 adopted for the size of the SF region (see Sect. 3.2).
 
 $^{c}$ Obtained at a poorer angular resolution ($\sim$ 20\arcsec) than the CO. 
 The peak column density could
 therefore be underestimated.
\end{table}

Kennicutt \cite{kennicutt98} used a  SFR a factor 1.5 higher and we adopt his value
when comparing to his data (Figure. 1, left panel).  We use a CO-to-molecular
gas conversion factor of $N_{\rm H2}/I_{\rm CO} = 2.0\times 10^{20}$ cm$^{-2}$ (K km s$^{-1})^{-1}$,
as in Bigiel et al. (2004) and Leroy et al. (2004),  but adopt to the
higher value of   Kennicutt \cite{kennicutt98}  of
$N_{\rm H2}/I_{\rm CO} = 2.8\times 10^{20}$ cm$^{-2}$(K km s$^{-1}) ^{-1}$
 in Figure 1 (left panel).

   \begin{figure}
   \centerline{
 \includegraphics[width=7.5cm,angle=270]{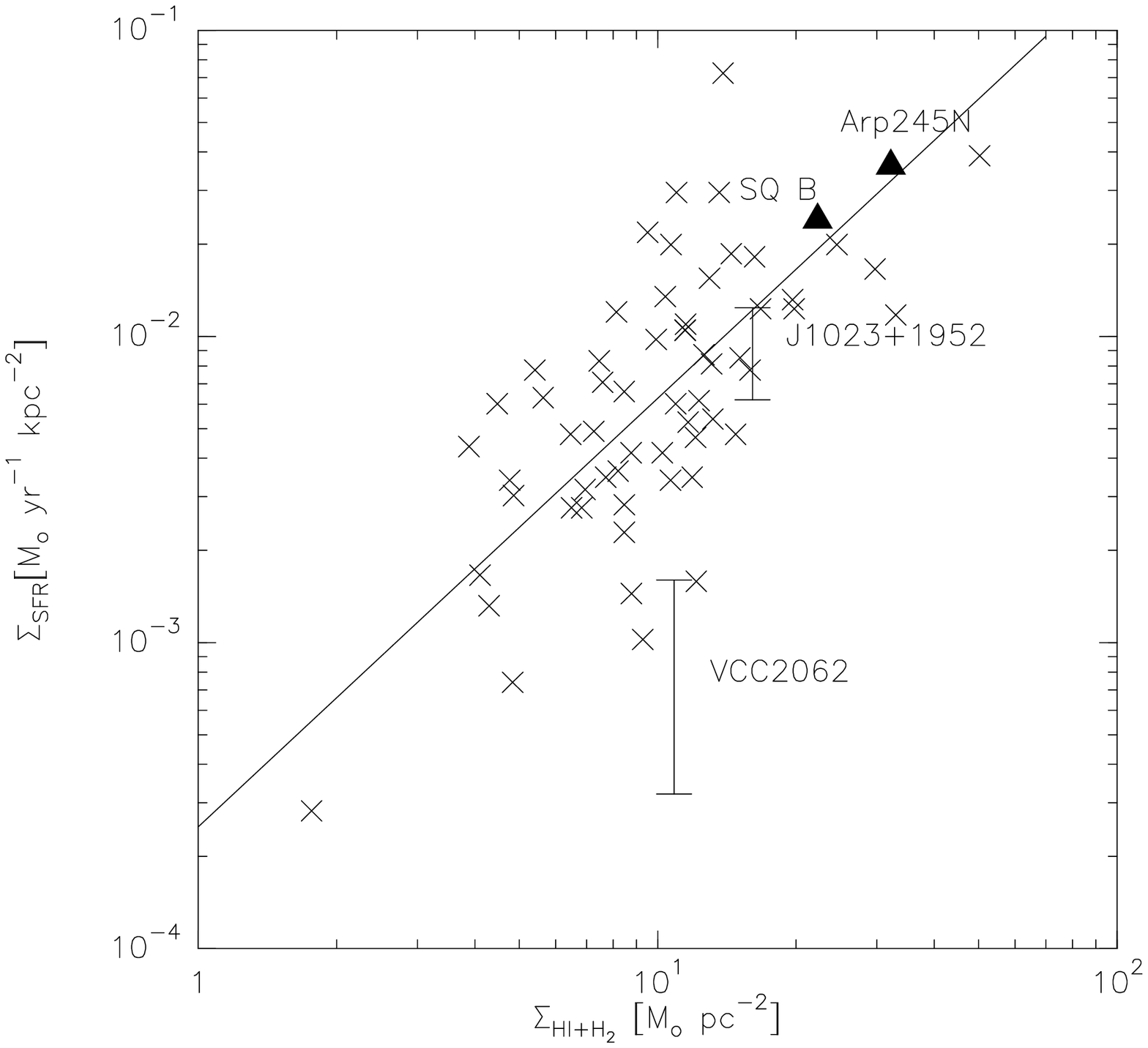}
 \includegraphics[width=7.5cm,angle=270]{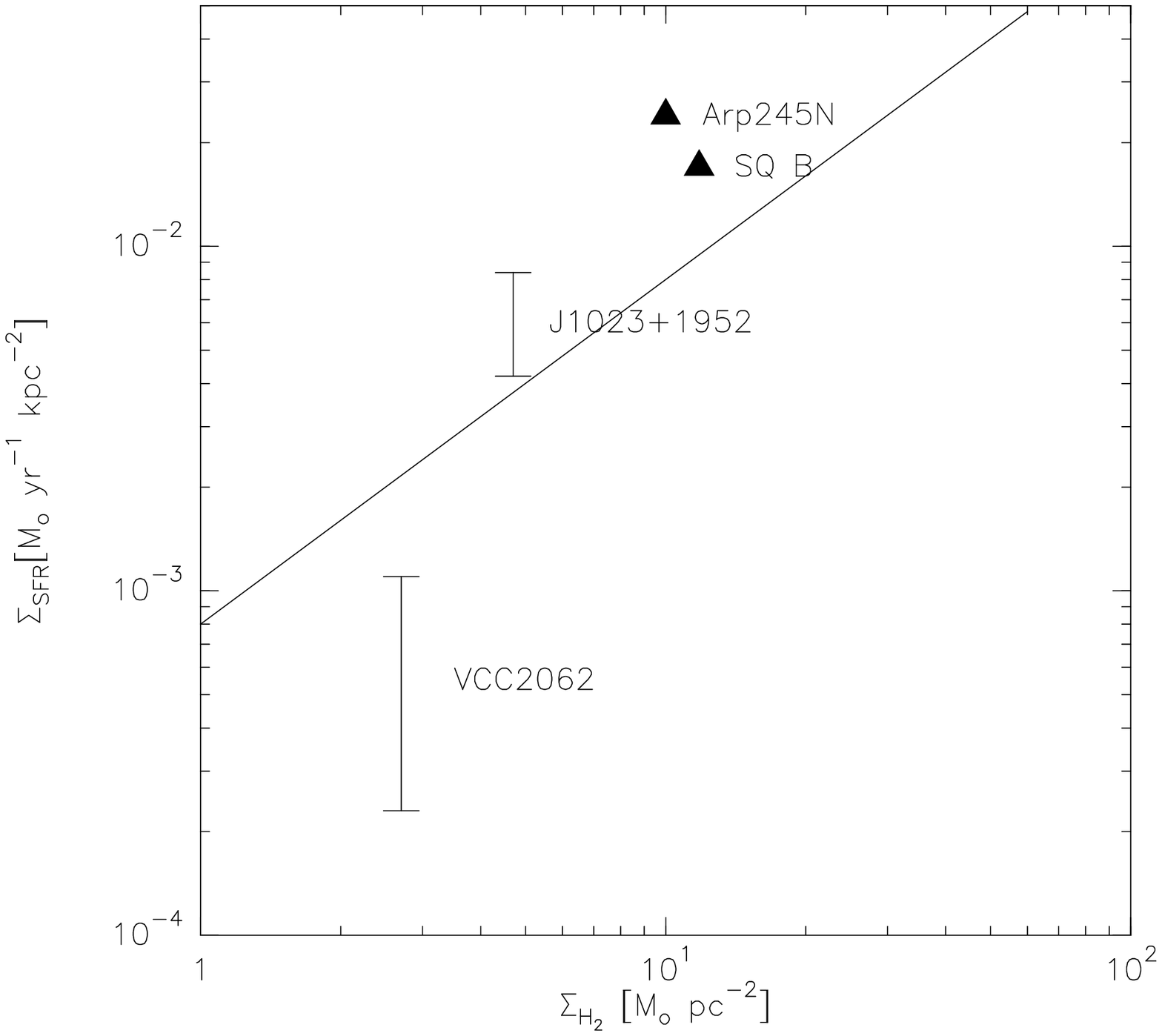}
 }
    \caption{{\it Left panel:} The relation between the SFR per area and the total gas mass per area 
   (the Schmidt law). The crosses are data 
    for a sample of spiral galaxies from Kennicutt \cite{kennicutt98}  (his Table  1) and the
    triangles are TDGs for which spatially resolved observations are available.
    The line gives the fit of Kennicutt \cite{kennicutt98}  to his entire sample consisting of
    spiral and starburst galaxies (his eq. 4).  We plotted our data in this figure after 
    correcting it for  his higher value of the SFR 
    and for his CO-to-molecular gas conversion factor (see text).
{\it Right panel:} The relation between the surface density of the SFR and of the molecular gas. The line gives the fit of Bigiel et al. \cite{bigiel} (their eq. 3) 
    to spatially resolved data of
    a sample of spiral galaxies.
    The range of  $\Sigma_{\rm SFR}$ given  for VCC2062 and J1023+1952,
    which only have CO data  at a modest resolution, is obtained  
    by  dividing the
    SFR by (i) the area of the SF region (upper end of the range) and (ii) by the area corresponding
    to the 22\arcsec\  CO beam (lower end).
    }
             \label{schmidtlaw}
   \end{figure}

In Table 1 we show the values  for the  molecular and  atomic gas surface densities,
$\Sigma_{\rm H2}$ and  $ \Sigma_{\rm HI}$, respectively,
and the SFR surface density, $ \Sigma_{\rm SFR}$.  
In the case of VCC2062 and J1023+1952 we give two values for
$ \Sigma_{\rm SFR}$. The lower value is calculated by
averaging the SFR over the area corresponding to the
the CO beam  in order to compare the data at the same angular resolution. 
The higher value gives the SFR averaged over the SF region only.
In Figure 1 we  compare these values to the Schmidt laws found by
Kennicutt \cite{kennicutt98} (left) and the molecular SF law found by Bigiel \cite{bigiel}.
Most objects  follow  both SF laws reasonably well, even
though $ \Sigma_{\rm SFR}$ of Arp245N lies a factor of 2 above the RMS scatter 
of 0.2 dex around the fit found for the THINGs galaxies \cite{bigiel}.
Only the data point of VCC 2062 lies well below the line, if $ \Sigma_{\rm SFR}$ 
is averaged over the area corresponding to the 22\arcsec\ resolution of the CO data.
The discrepancy disappears if $ \Sigma_{\rm SFR}$  is averaged over the SF region only.
Thus, if the local molecular gas surface density within the SF region is the same as the observed 
surface density averaged over the 22\arcsec\ beam, VCC 2062 follows the
Bigiel et al. relation. 
Higher resolution CO observations are necessary to resolve this issue.
A further uncertaintly comes from the use of the \halpha\ emission as a SF tracer.
Although we corrected the \halpha\ luminosity for extinction, dust-enshrouded SF
emitting very little in \halpha\ 
might be missed. In VCC 2062 this might be the case: the SFR derived from
the 8$\mu$m emission \cite{boquien}
 is about a factor of 4 higher than the SFR that we derived from the \halpha .
Thus, the additional use of a SF tracer sensitive to dust-enshrouded SF,
like the 25$\mu$m emission \cite{kennicutt07, leroy, bigiel}
is necessary to reliably trace the total SFR.

Leroy et al. \cite{leroy} found that the  molecular SFE is constant in spiral galaxies with
an average value of $(6\pm3) \times 10^{-10}$ yr$^{-1}$. They took into account a 
Helium fraction of 1.38 in the gas mass. The corresponding value without He,
suitable for comparison with our data, is SFE(H$_2) = (8\pm4) \times 10^{-10}$ yr$^{-1}$.  
Most of our values
 lie within this range, only the  value for Arp 245N lies  a factor of 2 above it.
 Leroy et al. \cite{leroy} also study the SFE in dwarf galaxies. Due to their low molecular
gas content, when adopting a Galactic conversion factor, they assumed that the gas
is dominated by the atomic hydrogen. They found in the central regions of dwarf galaxies
 much higher values for the SFE  than in the central regions of spiral galaxies.
The authors suggested that possibly an underestimate of the
molecular gas due to low metallicity  could be the reason. Our results give support to this
assumption because they show that (i) molecular gas is indeed present, although generally
not, as in the centers of spiral galaxies, in a dominant fraction and
(ii) the molecular SFE is  mostly in the same range as that  of  spirals. 

J1023+1952 \cite{lisenfeld08} and VCC 2062 \cite{duc07} are
particularly interesting objects to test the conditions necessary for the
onset of SF because spatially resolved observations are available
and both objects consist  of regions
with and without SF. We  can thus test the previously
explained criteria for the onset  of SF and search for differences 
 that can explain the presence or absence of SF.

In these two objects the total gas column density is high at all places of the gas cloud,
about 10 M$_\odot$ pc$^{-2}$
(VCC2062) and  10-20 M$_\odot$ pc$^{-2}$ (J1023+1952), respectively,
indicating that the minimum gas density to enable the presence of
cool gas is most likely
reached everywhere. 
%
In VCC2062, the position-velocity diagram (see Figure 6 in Duc et al. \cite{duc07}) allows us to 
 distinguish  clearly between the SF and non-SF region on a kinematical basis. The SF region is
 characterised by narrow lines and a velocity gradient indicating that
 the object is gravitationally bound. In the non-SF region no velocity gradient
is   visible and the lines are broad, suggesting that we are seeing HI left-over
 from the tidal tail which is not sufficiently dense to collapse to a bound object.

In J1023+1952 we were able to  estimate the values for the critical gas density according to
the Toomre criterion and according to the shear criterion   based on the velocity 
gradient of the HI gas presented in Mundell et al. \cite{mundell95} (their Figure 8) and
assuming that the velocity gradient is due  to
rotation.
We found for both criteria that the observed gas column density is
well above the critical value both in the SF and non-SF region of the gas clouds.
Thus, these criteria  cannot explain why SF is taking place in one
region and not in another.
Mundell et al. \cite{mundell04} and Lisenfeld et al. \cite{lisenfeld08} showed that SF in these
objects is dominated by a recent episode and they  found no indication
of an old stellar population. Therefore the gravitational potential of
old stars is not expected to play a role.

From the above we  conclude that the  line width
is also a relevant parameter in  explaining where SF occurs. 
 The spatial resolution  of the observation is 22\arcsec , corresponding to a 
linear scale of $\sim$ 2 kpc,  so that we do not resolve individual molecular clouds but
rather observe an ensemble of clouds within the beam. The CO 
line width therefore does not give us information about the
 properties of individual molecular clouds, but rather tells
 tells us something about the velocity dispersion
of the different molecular clouds that we observe within the beam.
A broad line width indicates that the velocity dispersion among the clouds is high,
i.e., it indicates large-scale turbulence.
The  low velocity dispersion observed in the SF region  seems  to be necessary for
SF to happen. We  can speculate that (i) the high velocity dispersion
in the rest of the gas cloud inhibits SF due to more frequent and more violent
cloud collisions or that (ii) the high turbulence goes in hand with a higher
scale height and thus a lower volume density of the cloud distribution making
collisions between them, and thus cloud collapse and ensuing SF, less likely.

\section{Summary and conclusions}

We reviewed observations of the molecular gas in TDGs and compared their
properties (column density, line width) to the SFR in order 
 to test, in a different environment,  the laws of SF derived for spiral and dwarf galaxies.
 In particular, we tested the Schmidt law by comparing the 
 column densities of the gas to the surface density
 of the SFR, derived from the extinction corrected H$\alpha$ luminosity,
  in four objects where CO and HI maps were available.
We compared the results of our TDGs to those of
spiral and starburst galaxies \cite{kennicutt98} and  of spiral galaxies 
from the THINGS sample \cite{bigiel}. We found a reasonable agreement 
with the spiral galaxies,  but noted in the case of VCC2062 the need for higher
resolution molecular gas data in order to correctly place the galaxy on the
correlation. 
Furthermore, we noted the need for using a SF tracer not affected
by extinction in order to trace also dust-enshrouded SF.

As far as the conditions for the onset of SF are concerned, we found
in two objects that a high column density of molecular gas is not  sufficient.
The line width is an additional important parameter to be taken into account
and SF only happens in those areas with
 narrow lines, indicating that a low velocity dispersion among the
molecular clouds is necessary for SF to take place.

\ack
This work has been supported by the Spanish Ministry of Science via the research
grants AYA 2005-07516-C02-01 and ESP 2004-06870-C02-02 and by the Junta de Andaluc\'\i a.

\end{document}